# Electrical control of hybrid exciton transport in a van der Waals heterostructure

Fedele Tagarelli[1,2*], Edoardo Lopriore[1,2*], Daniel Erkensten[3], Raül Perea-Causín[3], Samuel Brem[4], Joakim Hagel[3], Zhe Sun[1,2], Gabriele Pasquale[1,2], Kenji Watanabe[5], Takashi Taniguchi[6], Ermin Malic[4,3§], Andras Kis[1,2§]

[1]*Institute of Electrical and Microengineering, École Polytechnique Fédérale de Lausanne (EPFL), CH-1015 Lausanne, Switzerland*
[2]*Institute of Materials Science and Engineering, École Polytechnique Fédérale de Lausanne (EPFL), CH-1015 Lausanne, Switzerland*
[3]*Department of Physics, Chalmers University of Technology, 412 96 Gothenburg, Sweden*
[4]*Department of Physics, Philipps-Universität Marburg, 35037 Marburg, Germany*
[5]*Research Center for Functional Materials, National Institute for Materials Science, 1-1 Namiki, Tsukuba 305-0044, Japan*
[6]*International Center for Materials Nanoarchitectonics, National Institute for Materials Science, 1-1 Namiki, Tsukuba 305-0044, Japan*

*\* These authors contributed equally to this work.*
§*Correspondence should be addressed to: Andras Kis (andras.kis@epfl.ch) and Ermin Malic (ermin.malic@uni-marburg.de)*

**ABSTRACT**

Interactions between out-of-plane dipoles in bosonic gases enable the long-range propagation of excitons. The lack of direct control over collective dipolar properties has hitherto limited the degrees of tunability and the microscopic understanding of exciton transport. In this work, we modulate the layer hybridization and interplay between many-body interactions of excitons in a van der Waals heterostructure with an applied vertical electric field. By performing spatiotemporally resolved measurements supported by microscopic theory, we uncover the dipole-dependent properties and transport of excitons with different degrees of hybridization. Moreover, we find constant emission quantum yields of the transporting species as a function of excitation power with dominating radiative decay mechanisms over nonradiative ones, a fundamental requirement for efficient excitonic devices. Our findings provide a complete picture of the many-body effects in the transport of dilute exciton gases and have crucial implications for the study of emerging states of matter, such as Bose-Einstein condensation, as well as for optoelectronic applications based on exciton propagation.

**INTRODUCTION**

Exciton transport has been indicated as a potential basis for realizing scaled optical interconnects and modulators in chip-scale optical processing systems[1]. Strongly-bound and long-lived propagating excitons can act as carriers of information within a semiconductor, a desirable prospect for photonic circuits[2,3]. In particular, spatially-indirect excitons can propagate with micrometer-scale diffusion lengths and can be controlled *via* the quantum-confined Stark effect, enabling the tuning of their potential energy by an applied vertical electric field[4,5]. Van der Waals heterostructures of two-dimensional (2D) materials have been used as platforms for the manipulation of spatially-indirect interlayer excitons (IXs)[6]. In particular, type-II transition metal dichalcogenide (TMDCs) heterostructures have been employed to realize excitonic devices[7] and circuits[8]. In particular, heterostructures devices showing room temperature switching of exciton currents[7], tunable valley-polarized emission[9] and micrometer-scale transport of polarized exciton currents[10] have been demonstrated.

The spatial separation of charges comprising IXs gives rise to fixed out-of-plane dipole moments[11,12]. Repulsive Coulomb interactions between exciton populations of out-of-plane dipolar ensembles induce anomalous transport dynamics which deviates from standard diffusive propagation, characteristic of bosonic gases[13,14,15]. Instead, excitons generated in single TMDC layers, also called intralayer excitons, are mainly influenced by quantum-mechanical exchange interactions[16]. The interplay between dipolar and exchange interactions is highly dependent on the out-of-plane exciton dipole length, and layer-hybridized states with intra- and interlayer components are expected to show variable effective out-of-plane dipole lenghts[17]. The ability to control the degree of exciton hybridization is highly desirable as a means to modify the concurrent many-body interactions and tune the anomalous diffusion of exciton ensembles. While layer-hybridized states in twisted moiré heterostructures and their Stark shift with applied electric fields have been previously investigated[18–20], such type-II heterostructures do not allow tuning exciton-exciton interactions because the dipole moment of the emitting species is intrinsically fixed by layer arrangement. Moreover, the moiré superlattice that is formed in stacked bilayers has been shown to induce periodic potential traps that dramatically reduce the effective diffusivity of out-of-plane excitons[21,22], making moiré-less structures preferable for the manipulation of long-range propagating dipolar gases in excitonic devices.

Here, we exploit concurrent intervalley transitions in natural $WSe_2$ homobilayers to control the layer hybridization of exciton states by applying a vertical electric field. $2H-WSe_2$ homobilayers are moiré-less structures that have been indicated as a natural platform for Bose-Einstein condensation of interlayer exciton states[23]. However, an in-depth study of the dynamics and the transport of layer-hybridized tunable exciton states in this platform is still lacking. Furthermore, we achieve electrostatic control over hybrid exciton (hIX) transport in a structure with no moiré potential by varying the interplay between Coulombic dipolar repulsions and attractive exchange interactions. Our work sheds light on the influence of dipole length and hybridization on long-range interlayer exciton transport and is supported by a microscopic theory. Moreover, we show that the propagating exciton species in this platform are characterized by a quantum yield that is constant in power with dominant radiative recombination channels, independently on the layer hybridization. The study of electrically tunable dipolar ensembles with constant quantum yield and micrometer-scale transport opens the way for efficient excitonic devices based on van der Waals heterostructures of two-dimensional materials.

## RESULTS

**Electrically tunable interlayer dipolar ensembles in a van der Waals homobilayer**

Our devices consist of fully hBN-encapsulated WSe$_2$ natural homobilayers, featuring a bottom Cr/Pt gate and a top semitransparent Pt gate. The heterostacks are assembled on an SiO$_2$ substrate by mechanical transfers (see Methods). Figure 1c shows the image of an ultra-clean encapsulated bilayer WSe$_2$ device A, acquired by tapping mode Atomic Force Microscopy. Images of a second device (device B) are included in Supplementary Note 1. Natural 2H-WSe$_2$ homobilayers host momentum-indirect spin-bright $K\Lambda$ and $K\Lambda'$ transitions as the energetically lowest states[24,25] involving holes at the $K/K'$ points and electrons at the $\Lambda/\Lambda'$ points of the Brillouin zone (Figure 1a). Given their indirect nature, excited states in WSe$_2$ appear in the photoluminescence spectrum by their phonon replicas[24]. Intervalley excitons in bilayer WSe$_2$ are hybrid[26] in their intra- and interlayer components[27]. Our device architecture (Figure 1b) allows us to modulate the PL emission from layer-hybridized intervalley excitons in bilayer WSe$_2$ with an increasing applied vertical electric field $E_z$, causing the shift of the lowest-state transition from $K\Lambda$ ($K'\Lambda'$) to $K\Lambda'$ ($K'\Lambda$) (Table S1, Supplementary Note 2). The states in brackets represent degenerate states with opposite dipole moments[17]. In the presence of a positive vertical field with magnitude $E_z = 300$ mV/nm, the interlayer mixing coefficient of the energetically lowest state ($K\Lambda'$) is calculated to be $\left|C_{IX}^{K\Lambda'}\right|^2 = 0.80$.

The potential energy $U$ of out-of-plane electron-hole pairs with fixed dipole lengths $d$ can be modulated by the linear quantum-confined Stark effect as $\delta U \approx dE_z$. The degree of interlayer character of hIXs is highly tunable *via* an application of a vertical electric field. It can then be observed through the Stark effect acting on the out-of-plane component of the transitions of interest. From the field-dependent PL spectra in device A (Figure 1d-e), we distinguish two main ranges corresponding to the favorable transition being $K\Lambda$ and $K'\Lambda'$ ($|E_z| < 200$ mV nm$^{-1}$), or $K\Lambda'$ and $K'\Lambda$ ($|E_z| > 200$ mV nm$^{-1}$). From the linear Stark shift of the PL peaks with the highest intensity, we extract different effective out-of-plane dipole lengths $d_{\text{eff}}$ with respect to the vertical field based on the prevalent emitting states[28] (Figure 1f, Supplementary Note 3). In particular, $K\Lambda$ and $K'\Lambda'$ excitons are characterized by smaller dipole lengths ($d_{\text{eff}} \simeq 0.1$ nm) with respect to the $K\Lambda'$ and $K'\Lambda$ counterparts ($d_{\text{eff}} > 0.2$ nm)[28,29]. In our case, high positive and high negative vertical electric fields linked to $K\Lambda'$ and $K'\Lambda$ transitions are related to dipolar ensembles 0.41 nm and 0.24 nm long, respectively (Supplementary Note 3). Asymmetries in the field-dependent behavior of out-of-plane transitions have been previously reported as a function of doping[30]. In Supplementary Note 4 we show how the collective dipole moment of high-$d$ transitions can be effectively modulated by gating. In fact, effective out-of-plane dipole lengths of collective ensembles of optical excitations are tuned by induced or intrinsic doping within tens of angstroms due to electric field screening of the exciton wavefunctions[30]. Thus, we attribute the difference between the two branches in the Stark shift measurements of Figure 1e to the presence of intrinsic doping in the WSe$_2$ homobilayers used in this work.

**Field-effect control of hybrid exciton transport**

Our system hosts layer-hybridized excitons characterized by different effective lengths, allowing us to unveil the dipole-dependent transport properties of strongly interacting exciton gases in the dilute regime. Purely interlayer exciton gases with large separations between electrons and holes are characterized by negligible exchange forces[14,16,31]. On the other hand, hybrid ensembles with sizeable intra- and interlayer components host Coulombic dipolar repulsions and attractive exchange interactions[32]. By tuning the hybridization and the effective dipole length $d_{\text{eff}}$ of the probed excitons, we achieve control over the concurrent many-body interactions in the micrometer-scale transport of dilute exciton gases. We study tunable many-body interactions by measuring the steady-state effective diffusion area of hIXs in the presence of an applied electric field (Figure 2a). With negligible $E_z$, short-range exciton transport is observed due to the prevalence of the intralayer component in the energetically degenerate $K\Lambda$ and $K'\Lambda'$ states, featuring in this case randomly oriented dipoles and negligible repulsive interactions. However, with higher positive or negative fields, sizeable out-of-plane dipole lengths and larger interlayer mixing of hybrid states result in stronger collective repulsive forces. Consequently, we are able to electrostatically enhance the steady-state exciton gas expansion by a progressive transition from low-$d$ to high-$d$ dominating ensembles with increasing interlayer components (Figure 2b).

We can extract a lower-bound estimate of purely-interlayer exciton densities from the measured blueshift using the parallel-plate capacitor model[33]. However, layer-hybridized excitons in WSe$_2$ homobilayers show sizeable attractive exchange interactions resulting in a density-dependent redshift which counteracts the effect of dipolar repulsive Coulomb forces. In order to delve into the many-body picture of strongly-interacting dipolar gases, we have developed a microscopic theory accounting for the two main components driving excitonic transport in hybrid form. We study hIX interactions by deriving a hybrid exciton-exciton interaction Hamiltonian which we use to disentangle the main contributions to the density-dependent exciton energy renormalization (Supplementary Note 5):

$$\Delta E^{\xi}(n_x) = n_x^{\bar{\xi}} g_{d-d}^{\xi\bar{\xi}} + n_x^{\xi}\left(g_{d-d}^{\xi\xi} + g_{x-x}^{\xi\xi}\right) \qquad (1)$$

where $g_{d-d}$ is the dipole-dipole interaction strength, which is negligible for intralayer excitons in monolayer TMDCs and dominating for spatially separated interlayer excitons. Furthermore, $g_{x-x}$ accounts for exchange interactions, which are highly dependent on the out-of-plane separation for interlayer states[32]. The interactions are weighted by the valley-specific exciton density $n_x^{\xi}$, where the total exciton density $n_x$ is given by $n_x = \sum_{\xi} n_x^{\xi}$. Equation 1 takes into account all contributions from the different intervalley transitions, with $\xi = K\Lambda, K'\Lambda', K\Lambda', K'\Lambda$ ($\bar{\xi}$ denotes the opposite valley, i.e. if $\xi = K\Lambda$, then $\bar{\xi} = K'\Lambda'$). In Figure 2c-d, we show that the density-dependent energy normalization for hIXs in WSe$_2$ homobilayers is highly dependent on the vertical electric field as the hybrid exciton-exciton interaction crucially depends on the interlayer mixing coefficients. We have calculated the energy shifts for layer-hybridized excitons in undoped WSe$_2$ homobilayers by solving a hybrid exciton eigenvalue problem which accounts for mixing between intra- and interlayer exciton states[17], as described in Supplementary Note 2.

The effective dipole length $d_{\text{eff}}$ of the hIX is extracted by fitting a linear function to the energy shift $\Delta E^\xi = n_x d_{\text{eff}}^\xi / \epsilon$, with $\epsilon = \epsilon_0 \epsilon_\perp$ where $\epsilon_0$ is the vacuum permittivity and $\epsilon_\perp$ is the out-of-plane component of the dielectric tensor of the TMDC. The tunable effective out-of-plane dipole length of the exciton species is directly related to the level of layer hybridization, with $d_{IX} = 0.5$ nm for purely interlayer states. For the simulated ideal energy shifts for the dominant $K\Lambda'$ transition (interlayer mixing $\left|C_{IX}^{K\Lambda'}\right|^2 = 0.8$ for $E_z = 300$ mV/nm), we predict an effective dipole length $d_{\text{eff}} = 0.32$ nm, which is consistent with our measured values ranging from 0.24 nm to 0.41 nm in the presence of intrinsic doping. While excitons with a negligible interlayer component show no density-dependent shift in energy, as discussed in Supplementary Note 5, species with net sizeable out-of-plane dipole lengths show an increase in the blueshift magnitude at higher vertical electric fields. We characterize the measured renormalization shifts from high-$d$ (0.41 nm) and low-$d$ (0.24 nm) ensembles by observing a linear blueshift at low optical excitation intensity ($P_{in} < 150$ μW) followed by $\Delta E$ saturation. We ascribe this saturation to the presence of lattice heating at high exciton densities, as previously observed with spatially-indirect excitons in double quantum well systems (Supplementary Note 6).

**Radiative recombination of hIXs with a power-independent quantum yield**

Spatially-separated exciton species are characterized by longer radiative lifetimes with respect to their intralayer counterparts, since their electron and hole wavefunctions feature a smaller overlap and thus a lower probability to recombine[13]. Figure 3a shows the Stark shift of the main PL peaks and their measured lifetime in device B, highlighting the relationship between field-dependent lifetime and the change in the lowest-state emitting intervalley species. Comparable results are obtained in device A, as reported in Supplementary Note 7. Based on the longer effective out-of-plane dipole length of the hybrid excitations with higher fields, we observe an increase in lifetime when the main emitting transition shifts from $K\Lambda$ ($K'\Lambda'$) to $K\Lambda'$ ($K'\Lambda$). Excitons belonging to the former states in device B are characterized by average lifetimes of 0.65 ns, while the maximum value reached for the latter is around 0.75 ns (device A in Supplementary Note 7). Furthermore, we attribute the sudden drop in the hIX lifetime at large positive and negative fields to the dissociation of excitons and their tunneling through the hBN barriers to the top and bottom gate electrodes[34,35] (Supplementary Note 7).

It has been shown that the main nonradiative channel affecting the quantum yield of TMDCs is power-dependent exciton-exciton annihilation[36]. In our case, a single-exponential time-resolved PL decay is observed independently on the applied vertical electric field and on the input optical power (Supplementary Note 7). Moreover, a linear relationship between the integrated PL intensity and the input pump intensity indicates that the quantum yield of the probed exciton species is constant with power (Figure 3b). From the field-dependent data, we observe a linear increase in maximum PL intensity of hIXs with respect to lifetime (Figure 3c). A linear trend with positive coefficient, together with the constant quantum yield and single-exponential decays, indicate that the probed hIXs undergo mostly radiative recombination even at high excitation powers (Supplementary Note 7). These features draw a significant distinction between hIX dynamics and previously investigated purely-IX species in type-II band alignments, where density-dependent nonradiative terms have been shown to induce a decrease in the quantum yield of the dipolar ensembles at high excitation intensities[10,13,14].

**Time-resolved transport properties of tunable hIXs**

In order to fully understand the nature of interactions between propagating electrically tunable dipolar ensembles (Figure 4b), we study time-dependent hybrid exciton transport in our structures. To this purpose, we excite our sample with a picosecond pulsed laser and image the spatiotemporal expansion of the exciton cloud by its PL emission using a scanning avalanche photodiode system (Supplementary Note 8, Methods)[14,37]. The spatially-resolved exciton cloud corresponding to high-$d$ species is shown in Figure 4a for different points in time. The effective hIX area as a function of time is reported for high-$d$ and low-$d$ ensembles in Figure 4d. The equation of motion for the spatially resolved interlayer exciton density is derived as (Supplementary Note 9):

$$\partial_t n(\boldsymbol{r}, t) = D\nabla^2 n(\boldsymbol{r}, t) + \mu_m \nabla \cdot \left(\nabla\big(\Delta E(n(\boldsymbol{r}, t))\big) n(\boldsymbol{r}, t)\right) - \frac{n(\boldsymbol{r}, t)}{\tau} \quad (2)$$

which takes the form of a two-dimensional drift-diffusion equation, where $D$ is the diffusion coefficient, $\mu_m = D/k_B T$ the exciton mobility and $\tau$ the exciton lifetime. The energy renormalization term from Eq. 1 is now variable in space and time through $n(\boldsymbol{r}, t)$. Dipolar repulsions in the hIX equation of motion (Eq. 2) cause a nonlinear response to a pump excitation in the form of anomalous diffusion (Figure 4c). Thus, we introduce an effective diffusivity term $D_{\text{eff}}(t)$, defined as the slope of the exciton area of the hIX cloud distribution (Supplementary Note 9). Figures 4e-f show the time evolution of the simulated hIX area and effective diffusivity for the dipole lengths of interest. The initial exciton density in transport simulations is estimated based on a best-fit approach to the experimental results as $n_0 \simeq 10^{12}$ cm$^{-2}$ (Supplementary Note 9), below the exciton Mott transition limit in our system[38]. Since the highest exciton densities in the spot area are obtained for $n(\boldsymbol{r}, 0)$, the maximum hIX effective diffusivity is found at the limit $t \to 0$ for all dipolar species. Increasing $D_{\text{eff}}^{MAX}$ values are obtained for ensembles with higher $d_{\text{eff}}$, with $D_{0.24\text{nm}}^{MAX} \sim 7 \text{ cm}^2\text{s}^{-1}$ and $D_{0.41\text{nm}}^{MAX} \sim 11 \text{ cm}^2\text{s}^{-1}$ estimated from our simulations (Figure 4f). We note that the simulated transport for high-$d$ ensembles produces a faster initial anomalous diffusion regime with respect to the experimental data ($t < 1\text{ns}$), thus causing an overshoot of the resulting effective diffusivity. Instead, a good agreement is reached between theoretical and measured data in the low-$d$ case. Thus, given the trends observed in the anomalous diffusion regime, we conclude that both high-$d$ and low-$d$ species show similar maximum effective diffusivities in the range $5 - 10 \text{ cm}^2\text{s}^{-1}$, corresponding to an upper range of exciton mobility reaching $\mu_{\text{eff}}^{MAX} \sim 10000 \text{ cm}^2\text{V}^{-1}\text{s}^{-1}$ for high hIX densities and high-$d$ transitions in the regime of anomalous diffusion.

The effective diffusivities of all probed excitons decrease monotonically in time, progressively saturating to $D_{\text{eff}}(\infty) = D$, towards a regime of conventional diffusion. We note that $D_{\text{eff}}(\infty)$ is independent of the effective dipole length since it is equivalent to the conventional diffusivity for an exciton gas at low excitation densities. The unaltered $D$ is estimated experimentally by extracting the effective diffusivity of the hIX with minimal interlayer character. Thus, by measuring the propagation of exciton ensembles at $E_z = 0 \text{ mV/nm}$, we obtain a conventional diffusivity factor of $D \simeq 0.32 \text{ cm}^2\text{s}^{-1}$ (Supplementary Note 10).

**DISCUSSION**

Out-of-plane dipolar ensembles of optical excitations travel with long distances based on the strength of their repulsive forces, governed by the effective interlayer dipole length. In this work, we have achieved control over the layer hybridization of exciton states in a van der Waals homobilayer structure allowing us to tune the effective dipole length of exciton ensembles. We have characterized the dipole-dependent propagation of hIXs by modulating the interplay between attractive exchange interactions and repulsive Coulomb forces, the many-body effects governing exciton transport. Spatiotemporally resolved measurements have revealed a peak effective diffusivity of $\sim 10 \text{ cm}^2\text{s}^{-1}$, corresponding to an exciton mobility in the range $\mu_{\text{eff}}^{MAX} \sim 10000 \text{ cm}^2\text{V}^{-1}\text{s}^{-1}$.

The main factors affecting the efficiency of future interconnects and circuits based on exciton transport in van der Waals heterostructures are given by the material absorption, the exciton mobility, and the emission quantum yield. We have obtained power-independent high quantum yields of long-range propagating dipolar ensembles, a crucial step towards the efficient modulation of light in excitonic devices based on 2D materials.

Our microscopic understanding and control of the many-body effects governing the transport of dipolar exciton ensembles opens venues towards the exploration of exciton condensates in van der Waals structures and towards efficient excitonic devices based on two-dimensional materials.

**ACKNOWLEDGEMENTS**

This work was financially supported by the European Research Council (grant no. 682332) the Swiss National Science Foundation (grants no. 164015, 177007, 175822, 205114), and the Marie Curie Sklodowska ITN network "2-Exciting" (grant no. 956813). This project has received funding from the European Union's Horizon 2020 research and innovation programme under grant agreement No. 881603 (Graphene Flagship Core 3 Phase). Furthermore, the Marburg group acknowledges support from the Deutsche Forschungsgemeinschaft (DFG) vis SFB 1083.

## METHODS

**Device fabrication**

All the devices used in this work feature bottom gates fabricated by electron-beam lithography and metal evaporation (2nm Cr/5nm Pt) over a SiO$_2$/Si substrate with an oxide thickness of 270 nm. The heterostructures in devices B and C were fabricated with a polymer-assisted wet transfer method. WSe$_2$ (HQ Graphene) and hBN flakes were exfoliated on a polymer double layer, and bilayer WSe$_2$ flakes were identified by AFM measurements. The bottom polymer layer of the substrate was dissolved using a solvent, and the top polymer together with the exfoliated flakes was left free floating. The bottom hBN, WSe$_2$ bilayers and top hBN layers were then carefully aligned and transferred one over the other on the bottom gates by using a dedicated home-built transfer setup with motorized micromanipulators. On the other hand, the heterostructure of device A was fabricated using a dry-transfer technique employing polycarbonate (PC) membranes. WSe$_2$ (HQ Graphene) and hBN flakes were exfoliated on a SiO2 substrate, identified by optical contrast, and subsequently picked up using a single PC membrane. Then, the heterostack was released on the Cr/Pt bottom gate by progressive adhesion following a temperature gradient above 150°C. Such dry-transfer technique allowed us to obtain large area structures by avoiding contamination by polymer residues and water droplets. The optical images and AFM measurements of the devices are further shown in Supplementary Note 1. All the heterostructures were annealed in high-vacuum ($10^{-6}$ mbar) for six hours at a temperature of 340 °C. Finally, top gates and electrical contacts were fabricated by electron-beam lithography and evaporation of Pt (4nm) and Ti/Au (2nm/80nm) layers, respectively.

**Optical measurements**

All optical measurements were performed in vacuum at 4.6 K, unless stated otherwise, in a He-flow cryostat. Hybrid excitons are excited with a confocal microscope, meanwhile the emitted photons are collected through the same objective with a working distance of 4.5 mm and numerical aperture of 0.65. Optical pumping was achieved with a continuous-wave 640 nm diode laser (Picoquant, LDH-IB-640-M) focused to the diffraction limit (spot FWHM of 1.2 µm), for steady state measurements. For µPL spectral measurements, the emitted light was filtered by a 650 nm long-pass edge filter and then acquired using a spectrometer (Princeton Instruments SpectraPro 500) and recorded with a CCD camera (Princeton Instruments, Blaze 400-HR/HRX). Spatial imaging of the interlayer exciton emission was captured by a CCD camera (Andor Ixon) with an 800 nm long-pass edge filter that removes both the laser line and the intralayer emission from $WSe_2$. For time resolved measurement the same solid state laser is driven in pulsed mode, achieving pulse widths lower than 160 ps at 80 MHz repetition rate. The collected photons are sent to an APD (Excelitas Technologies, SPCM-AQRH-16) mounted on a 2D motorized translational stage. The output of the APD is connected to a time-correlated photon-counting module with a resolution of 12 ps r.m.s. (PicoQuant, PicoHarp 300), which measures the arrival time of each photon. For the measurements in this work, we set the time bin to 16 ps. The single-photon timing resolution of the APD is ~350 ps, which is the main time limitation for this setup. The technical details can be found in Supplementary Note 8.

**Microscopic many-particle theory**

In order to study hybridised exciton states and anomalous exciton transport at elevated electron-hole densities in TMD bilayers, we derive a many-body Hamiltonian in the hybrid exciton basis containing a kinetic part and a part due to exciton-exciton interactions. By solving the bilayer Wannier equation we get access to pure intra- and interlayer exciton states. These are used as input to a hybrid exciton eigenvalue problem which accounts for mixing between intra- and interlayer exciton states. We find that momentum-dark $K\Lambda$ ($K'\Lambda'$) excitons represent the energetically lowest-lying exciton states in naturally stacked $WSe_2$ bilayers. We include an out-of-plane electric field by exploiting the Stark shift of the interlayer exciton resonance, allowing us to tune the exciton landscape in bilayers as a function of electrical field (Supplementary Note 2). The obtained hybrid exciton states are then used to compute density-dependent energy renormalizations obtained via the Heisenberg equation of motion (Supplementary Note 5). By promoting the exciton density to be spatially dependent, we find that the exciton-exciton interaction acts as a source to a drift term in a drift-diffusion equation. We gain access to the spatiotemporal dynamics of hybrid excitons by solving the drift-diffusion equation for hybrid excitons with different field-driven interlayer mixing corresponding to different effective dipole moment lengths (Supplementary Note 9).

## AUTHOR CONTRIBUTIONS

A.K. initiated and supervised the project. E.L. and F.T. fabricated the devices, assisted by G.P.. F.T. performed the optical measurements, assisted by E.L. and Z.S.. K.W. and T.T. grew the h-BN crystals. F.T. and E.L. analysed the experimental data with input from A.K.. D.E., R.P.C., S.B., J.H. and E.M. developed the microscopic theory on exciton hybridisation and transport. E.L, F.T. and A.K. wrote the manuscript with contributions from all authors.

## COMPETING FINANCIAL INTERESTS

The authors declare no competing financial interests.

## DATA AVAILABILITY

The data that support the findings of this study are available on Zenodo at doi: XXXX.

## REFERENCES

1. Baldo, M. & Stojanović, V. Excitonic interconnects. *Nat. Photonics* **3**, 558–560 (2009).
2. Butov, L. V. Excitonic devices. *Superlattices Microstruct.* **108**, 2–26 (2017).

3. Perea-Causin, R. *et al.* Exciton optics, dynamics and transport in atomically thin semiconductors. Preprint at https://doi.org/10.48550/arXiv.2209.09533 (2022).

4. Grosso, G. *et al.* Excitonic switches operating at around 100 K. *Nat. Photonics* **3**, 577–580 (2009).

5. Shanks, D. N. *et al.* Interlayer Exciton Diode and Transistor. *Nano Lett.* **22**, 6599–6605 (2022).

6. Ciarrocchi, A., Tagarelli, F., Avsar, A. & Kis, A. Excitonic devices with van der Waals heterostructures: valleytronics meets twistronics. *Nat. Rev. Mater.* (2022) doi:10.1038/s41578-021-00408-7.

7. Unuchek, D. *et al.* Room-temperature electrical control of exciton flux in a van der Waals heterostructure. *Nature* **560**, 340–344 (2018).

8. Liu, Y. *et al.* Electrically controllable router of interlayer excitons. *Sci. Adv.* **6**, eaba1830 (2020).

9. Ciarrocchi, A. *et al.* Polarization switching and electrical control of interlayer excitons in two-dimensional van der Waals heterostructures. *Nat. Photonics* **13**, 131–136 (2019).

10. Unuchek, D. *et al.* Valley-polarized exciton currents in a van der Waals heterostructure. *Nat. Nanotechnol.* **14**, 1104–1109 (2019).

11. Rivera, P. *et al.* Observation of long-lived interlayer excitons in monolayer $MoSe_2$–$WSe_2$ heterostructures. *Nat. Commun.* **6**, 6242 (2015).

12. Merkl, P. *et al.* Ultrafast transition between exciton phases in van der Waals heterostructures. *Nat. Mater.* **18**, 691–696 (2019).

13. Jauregui, L. A. *et al.* Electrical control of interlayer exciton dynamics in atomically thin heterostructures. *Science* **366**, 870–875 (2019).

14. Sun, Z. *et al.* Excitonic transport driven by repulsive dipolar interaction in a van der Waals heterostructure. *Nat. Photonics* **16**, 79–85 (2022).

15. Lopriore, E., G. Marin, E. & Fiori, G. An ultrafast photodetector driven by interlayer exciton dissociation in a van der Waals heterostructure. *Nanoscale Horiz.* **7**, 41–50 (2022).

16. Erkensten, D., Brem, S. & Malic, E. Exciton-exciton interaction in transition metal dichalcogenide monolayers and van der Waals heterostructures. *Phys. Rev. B* **103**, 045426 (2021).

17. Hagel, J., Brem, S. & Malic, E. Electrical tuning of moir\'e excitons in MoSe$_2$ bilayers. Preprint at http://arxiv.org/abs/2207.01890 (2022).

18. Shimazaki, Y. *et al.* Strongly correlated electrons and hybrid excitons in a moiré heterostructure. *Nature* (2020) doi:10.1038/s41586-020-2191-2.

19. Ruiz-Tijerina, D. A. & Fal'ko, V. I. Interlayer hybridization and moiré superlattice minibands for electrons and excitons in heterobilayers of transition-metal dichalcogenides. *Phys. Rev. B* **99**, 125424 (2019).

20. Tang, Y. *et al.* Tuning layer-hybridized moiré excitons by the quantum-confined Stark effect. *Nat. Nanotechnol.* **16**, 52–57 (2021).

21. Choi, J. *et al.* Moiré potential impedes interlayer exciton diffusion in van der Waals heterostructures. *Sci. Adv.* **6**, eaba8866 (2020).

22. Li, Z. *et al.* Interlayer Exciton Transport in $MoSe_2WSe_2$ Heterostructures. *ACS Nano* **15**, 1539–1547 (2021).

23. Shi, Q. *et al.* Bilayer $WSe_2$ as a natural platform for interlayer exciton condensates in the strong coupling limit. *Nat. Nanotechnol.* **17**, 577–582 (2022).

24. Lindlau, J. *et al.* The role of momentum-dark excitons in the elementary optical response of bilayer $WSe_2$. *Nat. Commun.* **9**, 2586 (2018).

25. Deilmann, T. & Thygesen, K. S. Finite-momentum exciton landscape in mono- and bilayer transition metal dichalcogenides. *2D Mater.* **6**, 035003 (2019).

26. Wilson, N. R. *et al.* Determination of band offsets, hybridization, and exciton binding in 2D semiconductor heterostructures. *Sci. Adv.* **3**, e1601832 (2017).

27. Brem, S. *et al.* Hybridized intervalley moiré excitons and flat bands in twisted $WSe_2$ bilayers. *Nanoscale* **12**, 11088–11094 (2020).

28. Altaiary, M. M. *et al.* Electrically Switchable Intervalley Excitons with Strong Two-Phonon Scattering in Bilayer $WSe_2$. *Nano Lett.* **22**, 1829–1835 (2022).

29. Huang, Z. *et al.* Spatially indirect intervalley excitons in bilayer WSe$_2$. *Phys. Rev. B* **105**, L041409 (2022).

30. Wang, Z., Chiu, Y.-H., Honz, K., Mak, K. F. & Shan, J. Electrical Tuning of Interlayer Exciton Gases in WSe$_2$ Bilayers. *Nano Lett.* **18**, 137–143 (2018).

31. Woźniak, T., Faria Junior, P. E., Seifert, G., Chaves, A. & Kunstmann, J. Exciton g factors of van der Waals heterostructures from first-principles calculations. *Phys. Rev. B* **101**, 235408 (2020).

32. Kyriienko, O., Magnusson, E. B. & Shelykh, I. A. Spin dynamics of cold exciton condensates. *Phys. Rev. B* **86**, 115324 (2012).

33. Ivanov, A. L. Quantum diffusion of dipole-oriented indirect excitons in coupled quantum wells. *EPL Europhys. Lett.* **59**, 586 (2002).

34. Kash, J. A., Mendez, E. E. & Morkoç, H. Electric field induced decrease of photoluminescence lifetime in GaAs quantum wells. *Appl. Phys. Lett.* **46**, 173–175 (1985).

35. Sivalertporn, K., Mouchliadis, L., Ivanov, A. L., Philp, R. & Muljarov, E. A. Direct and indirect excitons in semiconductor coupled quantum wells in an applied electric field. *Phys. Rev. B* **85**, 045207 (2012).

36. Lien, D.-H. *et al.* Electrical suppression of all nonradiative recombination pathways in monolayer semiconductors. *Science* **364**, 468–471 (2019).

37. Akselrod, G. M. *et al.* Visualization of exciton transport in ordered and disordered molecular solids. *Nat. Commun.* **5**, 3646 (2014).

38. Siday, T. *et al.* Ultrafast Nanoscopy of High-Density Exciton Phases in WSe$_2$. *Nano Lett.* **22**, 2561–2568 (2022).

**FIGURES**

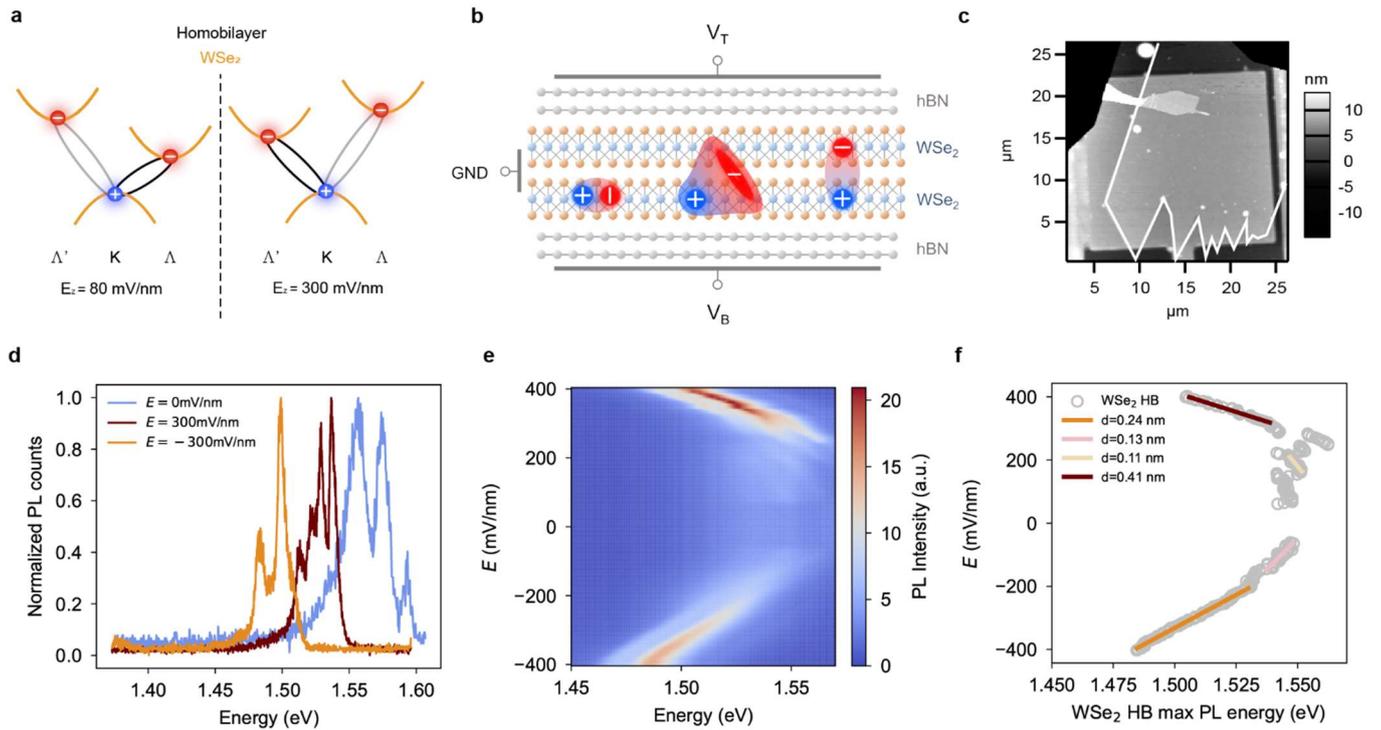

**Figure 1. Electrically tunable interlayer dipolar ensembles in a van der Waals homobilayer. a,** Schematic band structure of a natural WSe$_2$ homobilayer, hosting different dominant intervalley transitions based on the vertical electric field intensity (Supplementary Note 2). With a positive $E_z$, the main favorable transition shifts from $K\Lambda$ to $K\Lambda'$, with increasing interlayer mixing and sizeable out-of-plane dipole moments. **b,** Illustration of a double-gated fully hBN-encapsulated natural homobilayer WSe$_2$ device, with graphical representations of intralayer (left), hybrid (center) and purely interlayer (right) exciton species **c,** Atomic force microscopy (AFM) image of device A, with a large clean area ($> 80\ \mu m^2$) exhibiting uniform excitonic properties (Supplementary Note 1). **d,** Photoluminescence (PL) spectra acquired for different electric fields in device A, featuring the emission from hIX phonon replicas. **e,** PL spectra as a function of the applied vertical electric field. Low ($|E_z| < 200$ mV/nm) and high ($|E_z| \geq 200$ mV/nm) field regions are related to predominant $K\Lambda/K'\Lambda'$ and $K\Lambda'/K'\Lambda$ transitions, respectively. **f,** Extracted energy of the highest intensity PL peak from the PL spectra as a function of $E_z$. The energy shift as a function of electric field is fitted to a line in low and high-field regimes, revealing variable dipole moments. In particular, larger dipole lengths ($d > 0.2$ nm) are related to high-field regions dominated by $K\Lambda'$ and $K'\Lambda$ transitions, characterized by high interlayer mixing (Supplementary Note 2).

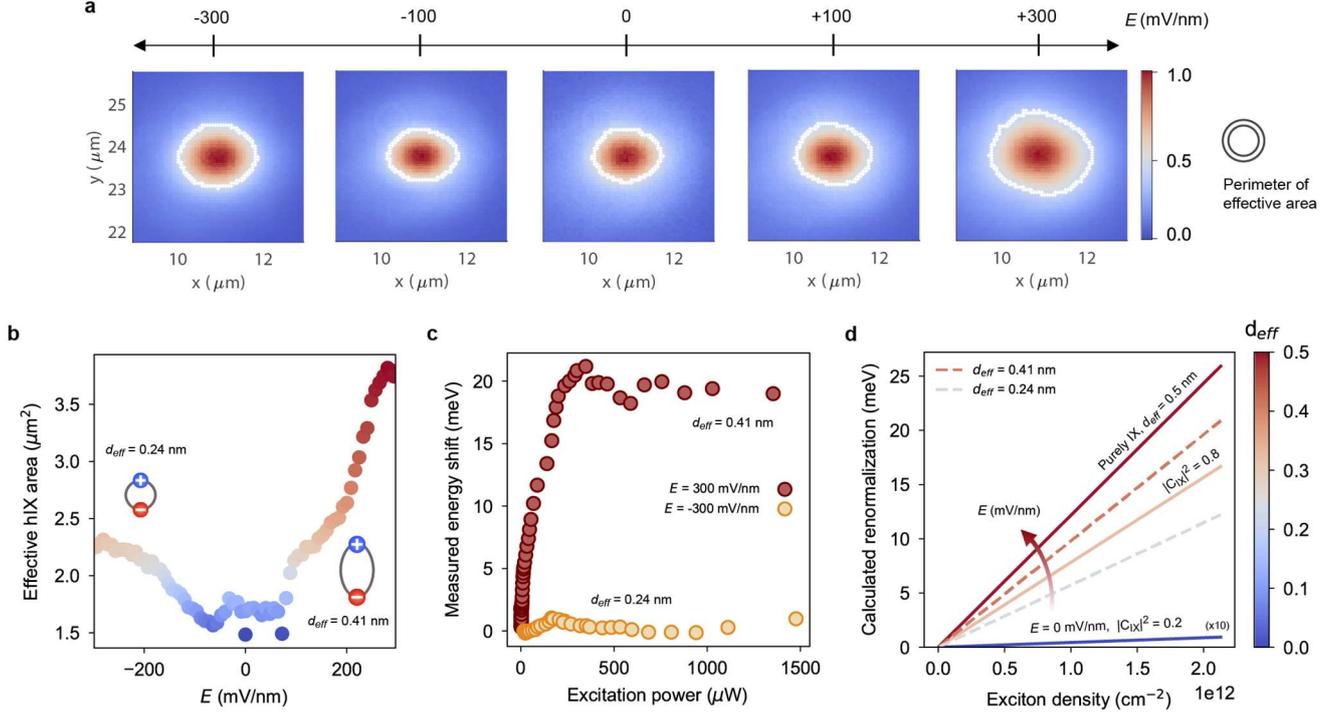

**Figure 2. Field-effect control of hybrid exciton transport. a,** Spatial images of steady-state hIX PL cloud expansion taken by a CCD camera (Methods) as a function of the applied vertical electric field. The perimeter of the effective area is defined as the 1/e points of the normalized PL intensity, and is shown as a white contour in all images for display purposes. **b,** Steady-state effective hIX cloud area plotted with respect to $E_z$. With increasing applied vertical electric fields, sizeable collective out-of-plane dipoles result in an increase in the repulsive interactions driving hIX diffusion and enhanced steady-state transport. The blue and red extremes in the colored representation correspond to negligible and high dipole lengths (right colorbar). The maximum areas obtained at positive and negative fields are related by a factor ~ 1.7 equivalent to the $d_{hd}/d_{ld}$ ratio between the corresponding dipole lengths of 0.41 nm and 0.24 nm, respectively. **c,** Extracted peak energy shift of the PL emission from the $K\Lambda'$ and $K'\Lambda$ transitions at high positive and negative electric fields as a function of the incident laser excitation power. Both species are characterized by linear blueshifts for $P_{in} < 150$ µW, followed by a saturation whose origin is discussed in detail in Supplementary Note 6. **d,** Theoretical calculation of the energy normalization shift in an ideal WSe$_2$ homobilayer as a function of the vertical electric field. Higher $E_z$ induce stronger interlayer mixing in the layer hybridization, with different linear energy shift tendencies and corresponding effective dipole lengths.

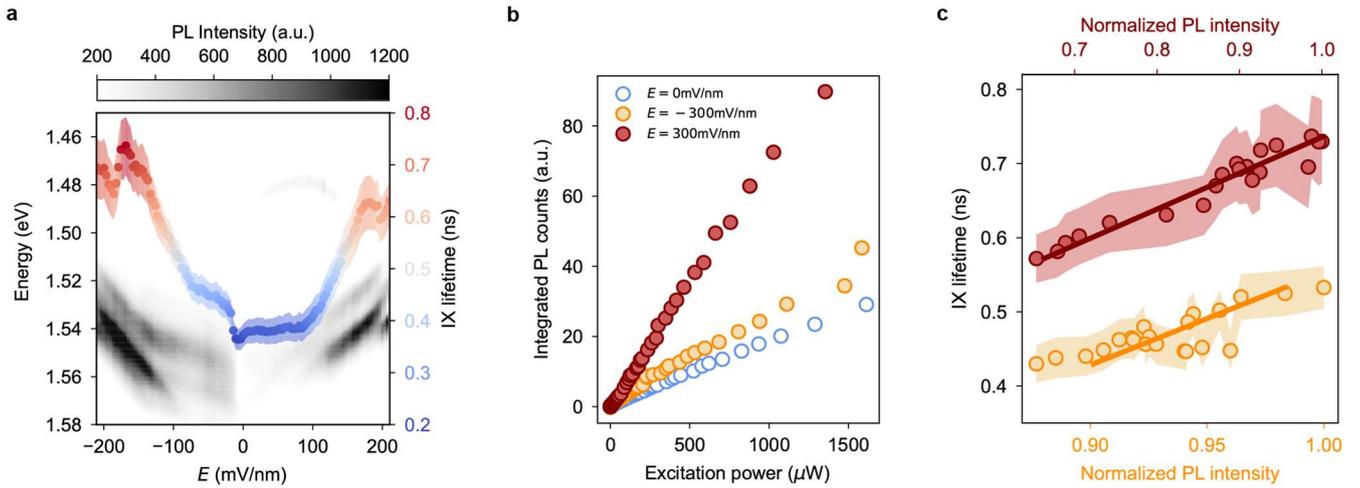

**Figure 3. Radiatively recombining hIXs with power-independent quantum yield. a,** The measured hIX lifetime in device B (color) is superimposed with the PL spectral map (black and white) as a function of the applied vertical electric field. The lifetime is extracted from a single-exponential fit to time-resolved PL measurements (Supplementary Note 7) and the calculated error is shown as a shaded area. The hIX lifetime increases with respect to the applied vertical electric field due to a smaller electron-hole wavefunction overlap at higher interlayer hybridizations. **b,** Integrated PL counts recorded at high positive, negative and zero electric fields in device A as function of laser excitation power. A constant emission quantum yield is recorded independently on $E_z$ and on the level of layer hybridization. The extended power regime in which PL counts are linear with respect to excitation power indicates that nonradiative decay mechanism are negligible in the probed hIX dynamics. **c,** A linear relationship is found between the hIX lifetime at high and low positive electric fields and their PL intensity, extracted from the data of Figure 3a. The $t_{hI} \propto I_{hIX}$ trend in our structure is independent on the layer hybridization and is explained by the predominance over radiative recombination channels in the transporting species (Supplementary Note 7).

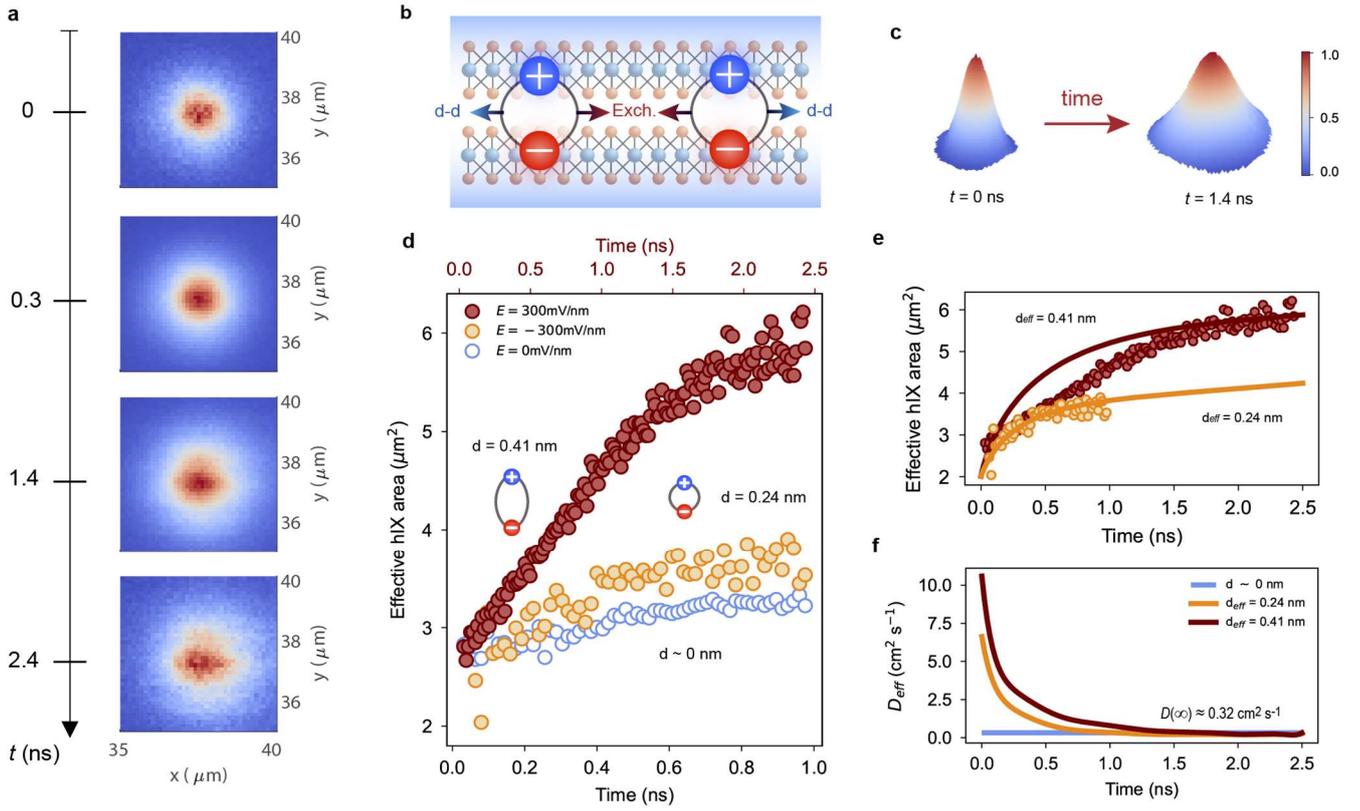

**Figure 4. Time-resolved transport properties of tunable hIXs. a,** Spatial imaging of the high-dipole hIX cloud expansion shown at different time delays $t$ with respect to the laser pulse arrival obtained at $E_z = +300$ mV/nm. **b,** Schematic representation of repulsive dipolar interactions and attractive exchange forces between out-of-plane ensembles in a van der Waals homobilayer. **c,** Three-dimensional representation of the measured hIX transporting cloud evolution from $t = 0$ ns to $t = 1.4$ ns. **d.** Extracted effective exciton diffusion area based on the 1/e threshold as a function of time measured at different electric fields. The extracted laser area with respect to time is plotted in Supplementary Note 8 as a reference. Anomalous diffusion is observed for hIXs at high positive and negative electric fields, while a linear expansion is obtained for negligible electric fields, indicating a classical transport for excitons with low interlayer mixing. **e,** Theoretically computed hybrid exciton diffusion area at different interlayer mixing, equivalent to different effective dipole moment lengths, computed using the exciton density $n_{hI}$ and the classical diffusivity factor $D$ extracted from experimental results. **f,** The effective diffusivity $D_{\text{eff}}$ is extracted from transport simulations with estimated maximum values of $D_{0.24\text{nm}}^{MAX} \sim 7$ cm$^2$s$^{-1}$ and $D_{0.41\text{nm}}^{MAX} \sim 11$ cm$^2$s$^{-1}$.